\documentclass[twocolumn]{aastex63}
\usepackage{amsmath}

\usepackage{lineno}

\newcommand{\subp}{_{\rm p}}

\newcommand{\subd}{_{\rm d}}

\newcommand{\OK}{\Omega_{\rm K}}

\renewcommand{\vec}[1]{\boldsymbol{#1}}

\newcommand{\be}{\begin{eqnarray}}
\newcommand{\ee}{\end{eqnarray}}

\newcommand\js{\bgroup\markoverwith{\textcolor[rgb]{0.8, .3, .1}{\rule[0.5ex]{8pt}{1.5pt}}}\ULon}

\shorttitle{Generating Highly-Eccentric Planets with Dispersing Eccentric Disks}
\shortauthors{Li and Lai}

\graphicspath{{./}{figure-use/}}

\begin{document}

\title{Resonant Excitation of Planetary Eccentricity due to a Dispersing Eccentric Protoplanetary Disk: 
A New Mechanism of Generating Large Planetary Eccentricities
}

\correspondingauthor{Jiaru Li}
\email{jiaru\textunderscore li@astro.cornell.edu}

\author[0000-0001-5550-7421]{Jiaru Li}
\affiliation{Center for Astrophysics and Planetary Science,
Department of Astronomy, Cornell University, Ithaca, NY 14853, USA}

\author[0000-0002-1934-6250]{Dong Lai}
\affiliation{Center for Astrophysics and Planetary Science, Department of Astronomy, Cornell University, Ithaca, NY 14853, USA}

\begin{abstract}
We present a new mechanism of generating large planetary eccentricities.
This mechanism applies to planets within the inner cavities of their companion protoplanetary disks.
A massive disk with an inner truncation may become eccentric due to non-adiabatic effects associated with gas cooling, and can retain its eccentricity in long-lived coherently-precessing eccentric modes;
as the disk disperses, the inner planet will encounter a secular resonance with the eccentric disk when the planet and the disk have the same apsidal precession rates;
the eccentricity of the planet is then excited to a large value as the system goes through the resonance.
In this work, we solve the eccentric modes of a model disk for a wide range of masses. 
We then adopt an approximate secular dynamics model to calculate the long-term evolution of the ``planet + dispersing disk'' system.
The planet attains a large eccentricity (between 0.1 and 0.6) in our calculations, even though the disk eccentricity is quite small ($\lesssim0.05$).
This eccentricity excitation can be understood in terms of the mode conversion (``avoided crossing'') phenomenon associated with the evolution of the ``planet + disk'' eccentricity eigenstates.
\end{abstract}

\keywords{accretion, accretion disks
-- hydrodynamics
-- celestial mechanics
-- planet-disk interactions
-- protoplanetary disks
}

\section{Introduction}  
\label{sec:intro}

Extrasolar planets (especially giant planets) are often observed to have non-negligible orbital eccentricities.
These eccentricities are usually interpreted as the results of planet-planet scatterings \citep[e.g.,][]{Rasio1996Sci,Weidenschilling1996Nature,Lin1997ApJ,Chatterjee2008ApJ,Juric2008ApJ,Frelikh2019ApJ,Anderson2020MNRAS,LJR2021MNRAS}, secular perturbations due to exterior stellar/planetary companions either through Lidov-Kozai oscillations \citep[e.g.,][]{Holman1997Nature,Wu2003ApJ,Fabrycky2007ApJ} or related co-planar effects \citep{Li2014ApJ,Anderson2017MNRAS}, or
planet-disk interactions.

In the conventional picture of planet-disk interactions, planetary eccentricity damping and growth are driven by resonant (Lindblad and corotatoin) torques.
For low-mass planets that do not open gaps in the disk, the co-orbital Lindblad torques dominate and lead to eccentricity damping \citep{Ward1986Icar,Artymowicz1993ApJ,Cresswell2007A&A}.
Recent studies suggest that thermal back-reaction effects (associated with the heat released by the planet and heat diffusion in the planet's near vicinity) can modify the torques significantly and lead to small (0.01-0.1) planetary eccentricities \citep{Eklund2017MNRAS,VelascoRomero2022MNRAS}.
For high-mass gap-opening planets, Lindblad torques can excite eccentricities, but corotation torques may damp them at a slightly faster rate \citep{Goldreich1980ApJ,Goldreich1981ApJ}.
Therefore, it is thought that planetary eccentricity growth may only occur for gap-opening planets with some (small) finite initial eccentricities that can saturate the corotation torques \citep{Goldreich2003ApJ,Ogilvie2003ApJ}.
On the other hand, an initially circular disk can become eccentric when the planet-to-star mass ratio is larger than about 0.003 \citep[e.g.,][]{Kley2006A&A,Regaly2010A&A,Teyssandier2017MNRAS}.
\cite{Teyssandier2016MNRAS} examined the back-reaction of the planet on the disk eccentricity and showed that when the dynamics of the eccentric disk is considered, the growth of planetary eccentricity is possible even when the corotation torques are not saturated. \cite{Ragusa2018MNRAS} carried out long-term ($>10^5$ orbits) hydrodynamical simulations of massive planets embedded in disks and showed that the planet can attain a small eccentricity ($\sim 0.15$).

This paper presents a new mechanism of generating large planetary eccentricities ($e\gtrsim0.5$) via secular planet-disk interactions.
This mechanism works as follows.
Consider a planet and a massive circumstellar disk orbiting around a central star.
The planetary orbit (initially circular) is co-planar with the disk and is confined inside the inner cavity of the disk.
Recent high-resolution observations show that many protoplanetary disks have inner cavities of various sizes \citep[e.g.,][]{vanderMarel2018ApJ,Huang2018ApJ,Long2018ApJ,Cieza2019MNRAS,Cieza2021MNRAS}; 
some are also found with inner planetary companions, non-axisymmetric substructures in the disks, and potential evidence of ongoing planet-disk interactions \citep[e.g.,][]{vanderMarel2013Sci,Keppler2018A&A,Long2022ApJ}.
\cite{LJR2021ApJ} have shown that a truncated and self-gravitating massive disk can become eccentric driven by non-adiabatic effects associated with disk cooling and can maintain its eccentricity in long-lived modes.
In this paper, we examine the long-term evolution of such an eccentric disk and its impact on the inner planet.
As the disk loses mass and disperses, the system crosses a secular apsidal resonance, which occurs when the disk and the planet have the same apsidal precession rates.\footnote{For other works that consider the secular interaction between planets and a dispersal outer disk, see also \cite{Petrovich2019AJ} and \cite{Teyssandier2019MNRAS}.}
The planet's orbital eccentricity is then excited to a large value after this resonance crossing.

The rest of this paper is organized as follows.
In Section~\ref{sec:Emode}, we present our model for eccentric protoplanetary disks.
We describe in Section~\ref{sec:planet-disk-interaction} the equations of the eccentricity evolution in our ``planet + outer disk'' system, and show in Section~\ref{sec:result} that the planet can acquire substantial eccentricity as the disk disperses. 
We explain the excitation of the planet eccentricities using a simple mode crossing analysis in Section~\ref{sec:AMD-mode}, and conclude in Section~\ref{sec:conclusion}.

\section{Disk Eccentric Mode and Instability}
\label{sec:Emode}

A protoplanetary disk is not perfectly axisymmetric in general.
Under certain conditions, a disk can be unstable due to small perturbations and develop a significant eccentricity.
\cite{LJR2021ApJ} carried out long-term hydrodynamical simulations of a massive self-gravitating disk with an inner cavity (and with a smoothly decaying outer density profile). 
They showed that because of gas cooling, such a disk naturally develops a coherent eccentric structure, which represents a global eccentric mode of the disk. 
This ``eccentric mode instability'' (EMI) arises because in the presence of gas cooling, the $m=1$ density perturbation and the corresponding pressure perturbation are out of phase, and the background disk exerts a net torque on the perturbed gas, which then amplifies the perturbation \citep[see also][]{Lin2011MNRAS,Lin2015MNRAS};  the self-gravity of the disk helps to turn these perturbations into a growing coherent eccentric mode.
The frequency, growth rate and radial profile of the eccentric mode can be calculated using a linear modal theory presented in \cite{LJR2021ApJ}.
For typical disk parameters, the mode growth time is about a few thousands disk rotation periods (at the inner disk edge).
\cite{LJR2021ApJ} also showed that the EMI saturates through non-linear hydrodynamical effects when the disk eccentricity reaches about 0.05; after saturation, the disk continues to evolve with a long-lived coherent eccentric pattern, which represents a \textit{stable} eccentric mode.

In the following we introduce the formalism that describes the disk eccentric modes. 
We further elaborate on the EMI and applying the formalism in Section~\ref{sec:planet-disk-interaction} and~\ref{sec:result} to calculate the secular planet-disk interaction.

An eccentric disk can be represented by a continuum of nested eccentric rings, with the complex eccentricity profile $E(r,t)$: all fluid elements at radius $r$ follow the same elliptical orbit with eccentricity $|E(r,t)|$ and longitude of pericenter $\arg\left[E(r,t)\right]$.
In a two-dimensional disk with small eccentricity, the time evolution of $E(r,t)$ is governed by a linear partial differential equation based on the fluid equations of mass and momentum conservations. 
Various forms of this evolution equation have been derived in the past to account for different equations of state, cooling and heating of the gas, and disk self-gravity \citep[e.g.,][]{Papaloizou2002A&A,Goodchild2006MNRAS,Teyssandier2016MNRAS,Lee2019ApJ,Lee2019ApJL,LJR2021ApJ}.
Schematically, the evolution of $E(r,t)$ reads 
\begin{eqnarray}
    \nonumber
    2r^3\OK\Sigma \frac{\partial E}{\partial t} = \bigg( &-&\frac{\beta}{i\beta+1} {\cal M}_{\rm adi} + \frac{i}{i\beta+1} {\cal M}_{\rm iso} \\
    \label{eq:E-equation-Ecool}
    & + & {\cal M}_{\rm sg} + {\cal M}_{\beta}\bigg)E,
\end{eqnarray}
where $\Omega_{\rm K}$ is Keplerian angular velocity of the disk, $\beta=\tau_{\rm c}\Omega_K$ is dimensionless cooling timescale \citep{Gammie2001ApJ}, and the four ${\cal M}$'s are the operators incorporating the effects of gas pressure, self-gravity and cooling.
The expressions for the ${\cal M}$'s can be found in Appendix A.3 of \cite{LJR2021ApJ}. 

An eccentricity eigenmode of the disk satisfies
\be
\label{eq:Emode-form}
    \partial E_{\rm m}/\partial t=i\omega_{\rm d,m}E_{\rm m},
\ee
where $\omega_{\rm d,m}$ is the (complex) eigenmode frequency.
While maintaining a constant radial shape, an eccentric mode experiences phase evolution coherently at all radii with the mode frequency $\omega_{\rm d,m}$.
That means the whole disk precesses like a rigid body when it is in an eccentric mode.

As a fiducial example, consider a system with a central star $M_{\star}$ surrounded by a protoplanetary disk with the density profile
\begin{eqnarray}
\nonumber
    \Sigma(r) = \Sigma_0
    &&\left[1+\tanh{\left(\frac{r-r_0}{w}\right)}\right]
    \left(\frac{r}{r_0}\right)^{-1/2} \\
\label{eq:disk-density}
    & &\times
    \exp{\left[-\left(\frac{r}{r_{\rm out}}\right)^{3/2}\right]},
\end{eqnarray}
where $r_0$ and $w$ are the location and the transition width of the disk's inner cavity, $r_{\rm out}$ is the outer radius of the disk, and $\Sigma_0$ is the surface density normalization that sets the total disk mass to $M_{\rm d}$.
Note that $\Sigma/\Sigma_0\ll1$ for $r-r_0\lesssim-2w$, thus $r_0$ characterizes the inner radius of the disk.
The disk sound speed profile is given by
\be
c_{\rm s}(r) = h_0 \left(\frac{GM_{\star}}{r_0}\right)^{1/2}\left(\frac{r}{r_0}\right)^{-1/4},
\ee
and the disk pressure is set to be $P(r) = c_{\rm s}^2 \Sigma/\gamma$ where $\gamma=3/2$ is the adiabatic index.
We adopt $r_{\rm out}=6r_0$, $w=0.1r_0$, $h_0=0.05$, and $\beta=10^{-6}$ for the fiducial values.

\begin{figure}[t]
    \epsscale{1}
    \plotone{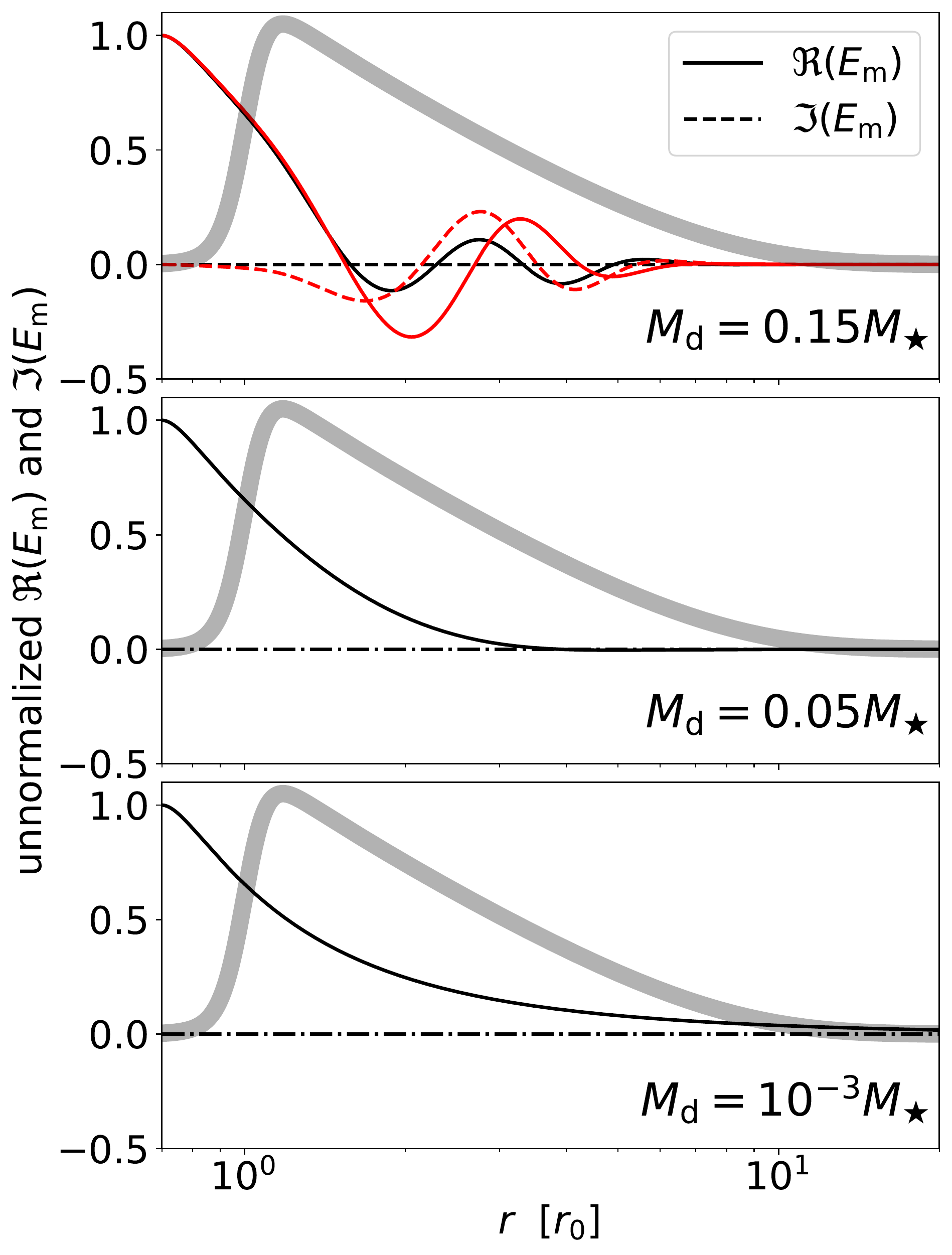}
    \caption{Fundamental eccentric modes of the fiducial disk. 
    The solid curves are the real parts of the $E_{\rm m}(r)$'s, while the dashed curves are the imaginary parts.
    The top panel shows the modes when $M_{\rm d}=0.15M_{\star}$, with the $e$-mode in black and the $s$-mode in red.
    The middle and the bottom panels are for $M_{\rm d}=0.05M_{\star}$ and $10^{-3}M_{\star}$; both cases allow only $e$-modes (black) but not $s$-modes.
    The $s$-mode has a frequency $\omega_{\rm d,m}=(0.012-0.001i)\Omega_{\rm K,0}$, where $\Omega_{\rm K,0} = \Omega_{\rm K}(r_0)$.
    The $e$-mode frequencies are $\omega_{\rm d,m}=0.010\Omega_{\rm K,0}$, $0.002\Omega_{\rm K,0}$, and $-0.0003\Omega_{\rm K,0}$ for $M_{\rm d}=0.15M_{\star}$, $0.05M_{\star}$, and $10^{-3}M_{\star}$, respectively.
    The disk surface density profile is shown as the grey curve for reference in each panel.
    }
    \label{fig:Emode-f}
\end{figure}

\begin{figure}[t]
    \epsscale{1.1}
    \plotone{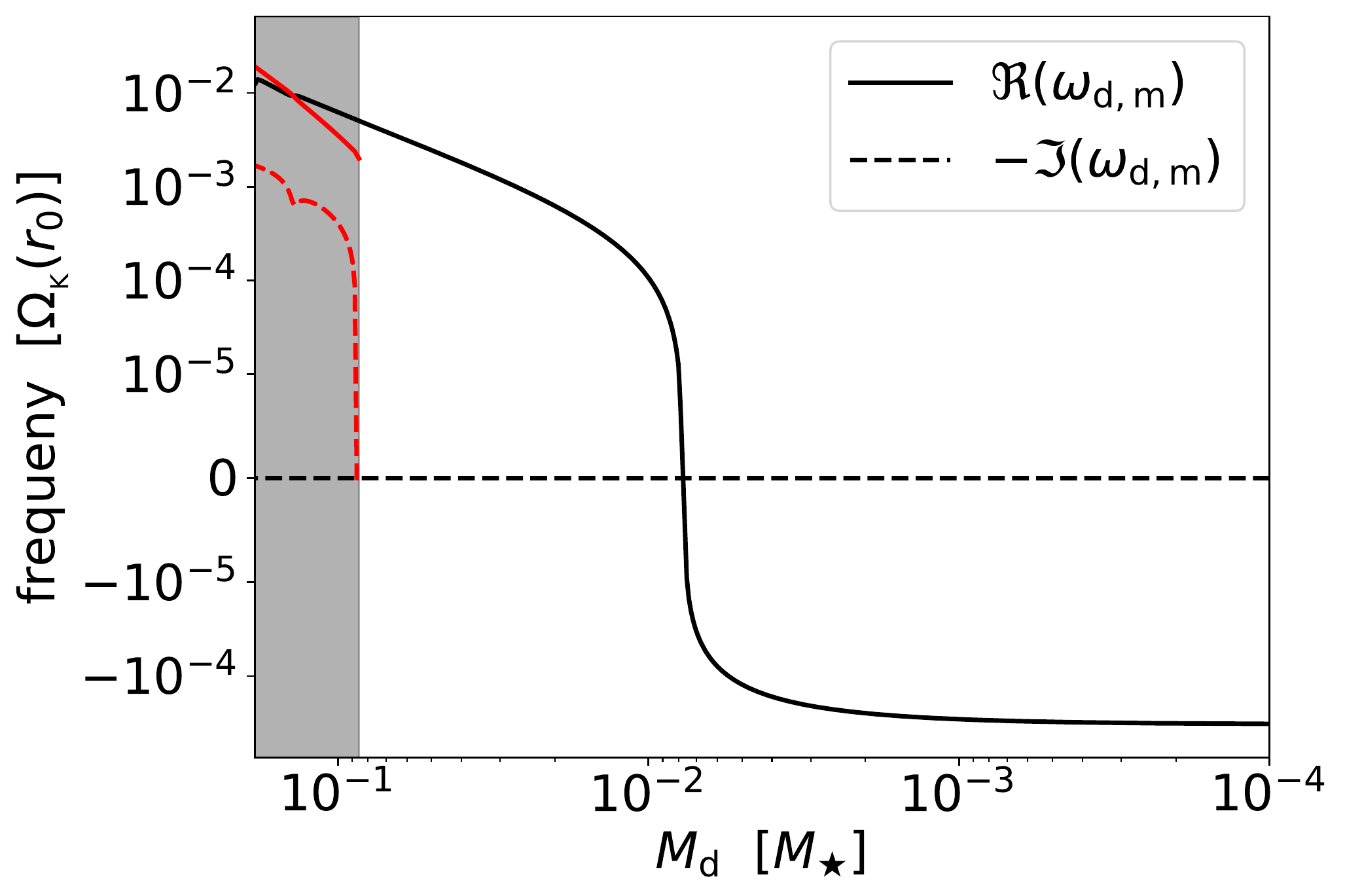}
    \caption{Frequencies of the fundamental eccentric modes of the fiducial disk as functions of $M_{\rm d}$. 
    The $e$-mode is shown in black and the $s$-mode is in red.
    The $s$-mode can only exist when the disk is massive (i.e., in the grey region).
    The solid curves show the real parts of the frequencies (in units of the Keplerian frequency at the disk inner radius), which represent the precession rates of the modes. 
    The dashed curves show the imaginary parts, corresponding to the growth rates of the modes.
    The $e$-mode always have $\Im{(\omega_{\rm d,m})}=0$.
    }
    \label{fig:Emode-w}
\end{figure}

Numerically solving Equations~\eqref{eq:E-equation-Ecool} and~\eqref{eq:Emode-form} for the fiducial disk finds two different kinds of eccentric modes: ``$e$-modes'' with real $E_{\rm m}$'s and ``$s$-modes'' with complex $E_{\rm m}$'s (with corresponding complex $\omega_{\rm d,m}$'s).
The disk eccentricities at different radii are apsidally aligned for an $e$-mode and not aligned for an $s$-mode.
Figure~\ref{fig:Emode-f} shows the radial profiles of the fundamental $e$ and $s$-modes (i.e., the mode with the least number of radial nodes) for disks with different $M_{\rm d}$'s.
When $M_{\rm d} = 0.15M_{\star}$, the disk possesses both $e$-modes and $s$-modes.
When $M_{\rm d} = 0.05M_{\star}$ and $10^{-3}M_{\star}$, the disk only supports $e$-modes. 
As $M_{\rm d}$ decreases, the $E_{\rm m}(r)$ profile for the $e$-mode evolves continuously.

Figure~\ref{fig:Emode-w} shows the mode frequency $\omega_{\rm d,m}$ as a function of $M_{\rm d}$ for our fiducial disk model.
The $e$-mode (black) exists at all $M_{\rm d}$'s.
Its frequency depends on $M_{\rm d}$:
in a massive disk, the precession of the $e$-mode is mainly driven by the disk's self-gravity, so the precession frequency $\omega_{\rm d,m}>0$, implying prograde apsidal precession of the disk;
in a low-mass disk with negligible self-gravity, the precession is due to the pressure force and $\omega_{\rm d,m}<0$. 
The transition from positive to negative allows the precession rate of a dispersing disk to sweep a wide range and cross a secular resonance with a companion planet (see Section~\ref{sec:result}).
The imaginary part $\Im{(\omega_{\rm d,m})}$ of the $e$-mode is always zero, implying that this mode is stable and can help the disk maintain long-lived eccentricity.

The $s$-mode (red in Figure~\ref{fig:Emode-w}) only exists when $M_{\rm d}$ is large.
It has an imaginary part $\Im{(\omega_{\rm d,m})}<0$, which means that the mode amplitude grows exponentially with time in the linear regime;
this eccentricity growth is the EMI.\footnote{Although EMI occurs in our fiducial disk only when $M_{\rm d}\gtrsim0.1M_{\star}$, it has also been found in much less massive disks with about $1\%$ stellar mass \citep[see Figures 7 and 9 in][]{LJR2021ApJ}.}
A massive disk can acquire a non-negligible eccentricity rapidly before the gas depletes through the growth of the $s$-mode, then transfer the eccentricity to an $e$-mode through non-linear effects when the growth saturates \citep[see][]{LJR2021ApJ}.

In summary, for a disk that is initially massive but loses mass with time, it can first become eccentric due to the $s$-mode instability, then maintain a long-lived coherent eccentric pattern due to the $e$-mode.

\section{Planet-Disk Interaction Model}
\label{sec:planet-disk-interaction}

We now model how a disk with a long-lived coherent eccentricity $E(r,t)$ (in the form of disk $e$-mode) can influence the eccentricity of a companion planet. 
We assume the planet (with nass $M_{\rm p}$ and semi-major axis $a_{\rm p}$) resides inside the inner cavity of the disk on a co-planar orbit; the planet's complex eccentricity is denoted by $E_{\rm p}(t)$.

We adopt the method described in \cite{Teyssandier2019MNRAS} to derive a simplified secular model for eccentricity evolution of the ``planet + disk'' system.
This method has been found to reproduce the key features of planet-disk secular interaction in hydrodynamical simulations.
First, we assume that the disk's coherent complex eccentricity $E(r,t; M_{\rm d})$ has the same ``shape'' as the fundamental $e$-mode of the disk.
This allows us to write $E$ as
\be
\label{eq:E-fE}
    E(r,t; M_{\rm d}) = E_{\rm m}(r; M_{\rm d}) E_d(t),
\ee
where $E_{\rm d}(t)$ specifies the complex amplitude of the mode, and $E_{\rm m}(r; M_{\rm d})$ is the mode eigenfunction discussed in Section~\ref{sec:Emode}.
Note that $E_{\rm m}(r)$ is real (for $e$-mode) and depends on $M_{\rm d}$, so we denote it by $E_{\rm m}(r; M_{\rm d})$.
Equation~\eqref{eq:E-fE} assumes that the disk precesses coherently with a common $E_{\rm d}(t)$ for all rings.
We normalize $E_{\rm m}(r; M_{\rm d})$ by
\begin{eqnarray}
    J_{\rm d} \equiv \int \Sigma r^2 \Omega_{\rm K} |E_{\rm m}|^2 2\pi r dr = \int \Sigma r^2 \Omega_{\rm K} 2\pi r dr.
\end{eqnarray}
Note that the angular momentum deficit of the disk is given by 
\begin{equation}
A_{\rm d} = \int \frac{1}{2}|E|^2\Sigma r^2 \Omega_{K} 2\pi rdr = \frac{1}{2} J_{\rm d} |E_{\rm d}|^2.
\end{equation}

Adopting Equation~\eqref{eq:E-fE} for the disk eccentricity, the secular evolution of the ``planet + disk'' system is fully described by the equations of motion for $E_{\rm d}(t)$ and $E_{\rm p}(t)$ \citep{Teyssandier2019MNRAS}:
\begin{eqnarray}
\label{eq:dEddt}
    \frac{dE\subd}{dt} 
    & = & i (\omega_{\rm d,m}+\omega_{\rm d,p}) E\subd - i \nu_{\rm d,p} E\subp, \\
\label{eq:dEpdt}
    \frac{dE\subp}{dt}
    & = & - i \nu_{\rm p,d} E\subd + i \omega_{\rm p,d} E\subp.
\end{eqnarray}
Here $\omega_{\rm d,m}$ is the ``intrinsic'' precession rate (driven by pressure and self-gravity) of the disk (Equation~\ref{eq:Emode-form}), and $\omega_{\rm d,p}$ and $\omega_{\rm p,d}$ are the precession rates of the disk and planet due to disk-planet interaction:
\begin{eqnarray}
\label{eq:wdp}
    \omega_{\rm d,p} 
    & = & \frac{1}{J\subd} \int G M\subp \Sigma K_1(r,a_{\rm p}) |E_{\rm m}|^2 2 \pi r dr, \\
\label{eq:wpd}
    \omega_{\rm p,d}
    & = & \frac{1}{J\subp} \int G M\subp \Sigma K_1(r,a_{\rm p}) 2 \pi r dr,
\end{eqnarray}
where $J_{\rm p}=M_{\rm p}a_{\rm p}^2\Omega_{p}=M_{\rm p}\sqrt{GM_{\star}a_{\rm p}}$ is the zero-eccentricity angular momentum of the planet.
In Equations~\eqref{eq:dEddt} and~\eqref{eq:dEpdt}, the coefficients $\nu_{\rm d,p}$ and $\nu_{\rm p,d}$ are the ``eccentricity coupling'' rates, given by
\begin{eqnarray}
\label{eq:vdp}
    \nu_{\rm d,p} 
    & = & \frac{1}{J\subd} \int G M\subp \Sigma K_2(r,a_{\rm p}) E_{\rm m} 2 \pi r dr, \\
\label{eq:vpd}
    \nu_{\rm p,d}
    & = & \frac{1}{J\subp} \int G M\subp \Sigma K_2(r,a_{\rm p}) E_{\rm m} 2 \pi r dr.
\end{eqnarray}
Note that $\nu_{\rm d,p} J_{\rm d} = \nu_{\rm p,d} J_{\rm p}$.
In Equations~\eqref{eq:wdp}-\eqref{eq:vpd}, the $K$'s are related to the Laplace coefficients of celestial mechanics and are defined in \cite{Teyssandier2016MNRAS}.

For reference, we note that for a disk with a very narrow width (so that $r_0\approx r_{\rm out}$), the various frequencies become 
\begin{eqnarray}
\label{eq:w-app}
    \omega_{\rm p,d} & = &\frac{J_{\rm d}}{J_{\rm p}}\omega_{d,p} = \frac{M_{\rm d}}{4M_{\star}} \alpha^2 \Omega_{\rm p} b_{3/2}^{(1)} (\alpha), \\
\label{eq:v-app}
    \nu_{\rm p,d} & = &\frac{J_{\rm d}}{J_{\rm p}}\nu_{d,p} = \frac{M_{\rm d}}{4M_{\star}} \alpha^2 \Omega_{\rm p} b_{3/2}^{(2)} (\alpha), 
\end{eqnarray}
where $\alpha=a_{\rm p}/r_0$, $b_{3/2}^{n}(\alpha)$ is the Laplace coefficient, and $b_{3/2}^{(1)}\simeq3\alpha$, $b_{3/2}^{(2)}(\alpha)\simeq15\alpha^2/4$ for $\alpha\ll1$.

\section{Planet Eccentricity Excitation: Numerical Examples}
\label{sec:result}

To understand the evolution of the planet eccentricity as the disk disperses, we integrate Equations~\eqref{eq:dEddt} and~\eqref{eq:dEpdt} starting from an initially eccentric disk with $M_{\rm d,0}=0.1M_{\star}=0.1M_{\odot}$.
We use the same disk model as in Section~\ref{sec:Emode}.
At $t=0$ (the starting point of the integration), the disk has already gone through a rapid phase of eccentricity growth via EMI, and we set the initial disk eccentricity amplitude to $E_{\rm d}=0.05$.
We assume that disk mass decreases in time as
\begin{eqnarray}
\label{eq:Mt}
    M_{\rm d}(t) = \frac{M_{\rm d,0}}{1+t/\tau_{\rm d}},
\end{eqnarray}
where $\tau_{\rm d} = 10^5P_0$ is the disk decay timescale with $P_0$ being the Keplerian orbital period at $r_0$.
The disk eccentric mode profile $E_{\rm m}(r;M_{\rm d})$ is set to be the same as the $e$-mode profile computed as in Section~\ref{sec:Emode}. 
A planet $M_{\rm p} = 3\times10^{-4}M_{\odot}$ is placed inside the disk cavity at $a_{\rm p}=0.2r_0$ and is assumed to have an initial $E_{\rm p}=0$.
We refer to the ``planet + disk'' system described above as the fiducial system.

\begin{figure}[t]
    \epsscale{1}
    \plotone{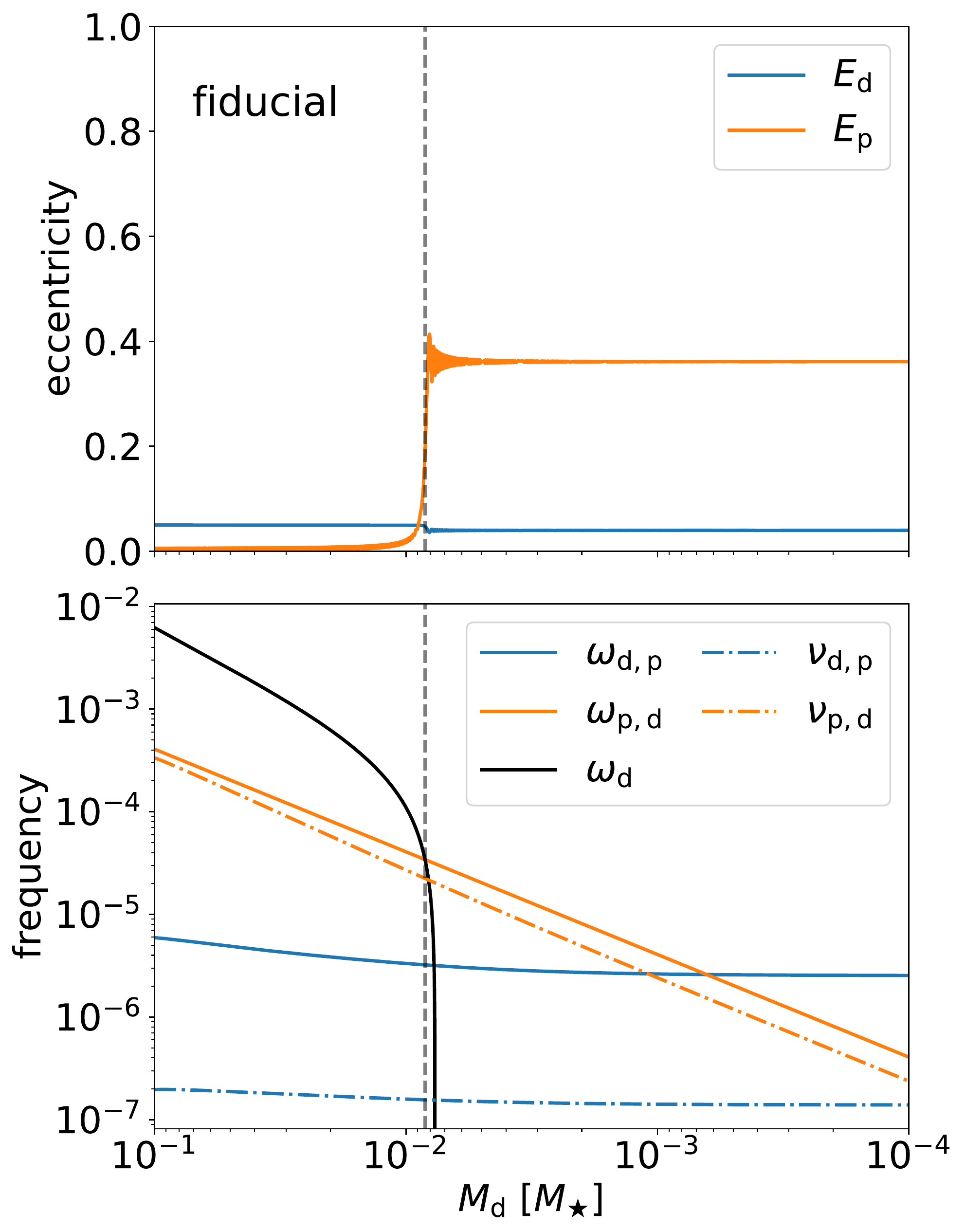}
    \caption{Result of the long-term time evolution in the the fiducial system ($a_{\rm p}=0.2r_0$, $M_{\rm p}=3\times10^{-4}M_{\odot}$, and $\tau_{\rm d}=10^{5}P_0$).
    The disk mass $M_{\rm d}$, which is used as the horizontal axes in both panels, is a function of time (see Equation~\ref{eq:Mt}).
    The upper panel shows the eccentricity evolution of the planet and the disk.
    The lower panel shows the behaviors of the precession rates and the eccentricity coupling rates for the secular dynamics model (Equations~\ref{eq:dEddt} and~\ref{eq:dEpdt}).
    The vertical dashed grey lines mark the $M_{\rm d}$ value when $\omega_{\rm d}=\omega_{\rm d,m}+\omega_{\rm d,p}=\omega_{\rm p,d}$.
    }
    \label{fig:res-fid}
\end{figure}

In Figure~\ref{fig:res-fid} we show the result of our fiducial evolution.
The lower panel shows the behaviors of $\omega_{\rm d,p}$ (solid blue), $\omega_{\rm p,d}$ (solid orange), $\nu_{\rm d,p}$ (dashed blue), $\nu_{\rm p,d}$ (dashed orange), and $\omega_{\rm d} = \omega_{\rm d,m} + \omega_{\rm d,p}$ (solid black) as the disk mass decreases.
The precession rate $\omega_{\rm p,d}$ and the coupling rate $\nu_{\rm p,d}$ are induced by the disk, so they diminish linearly with the disk mass. 
The forced precession rate of the disk $\omega_{\rm d,p}$ and the coupling rate $\nu_{\rm d,p}$ are induced by the planet, so they only slightly decrease as $E_{\rm m}(r;M_{\rm d})$ changes slowly.
The net precession rate of the disk $\omega_{\rm d} = \omega_{\rm d,m} + \omega_{\rm d,p}$ is dominated by $\omega_{\rm d,m}$ for very large and very small $M_{\rm d}$. 
As a result, $\omega_{\rm d}$ is much larger than the planet precession rate $\omega_{\rm p,d}$ initially, becomes equal to $\omega_{\rm p,d}$ as the disk mass decreases, and converges to a negative value at small disk mass after the disk self-gravity becomes negligible (see Figures~\ref{fig:Emode-w} and~\ref{fig:res-fid}).

The upper panel of Figure~\ref{fig:res-fid} shows the eccentricity evolution.
We see that $E_{\rm d}$ and $E_{\rm p}$ are nearly constant at the beginning and the end.
However, when $\omega_{\rm d} \simeq \omega_{\rm p,d}$, the planet eccentricity is quickly excited to $E_{\rm p}=0.36$, and the disk eccentricity $E_{\rm d}$ decreases correspondingly by a small amount.
This means the planet-disk eccentricity exchange is the strongest when the disk eccentricity precession and the planetary apsidal precession are in resonance.
As the disk dissipates and $\omega_{\rm d}$ sweeps from positive to negative, this resonance only occurs once and generates a one-time fast eccentricity transfer.
At the resonance, $M_{\rm d}\sim 10^{-2}M_{\star}\gg M_{\rm p}$, so the planet can receive a final $E_{\rm p}$ that is much higher than the initial $E_{\rm d}$.

\begin{figure}[t]
    \epsscale{1}
    \plotone{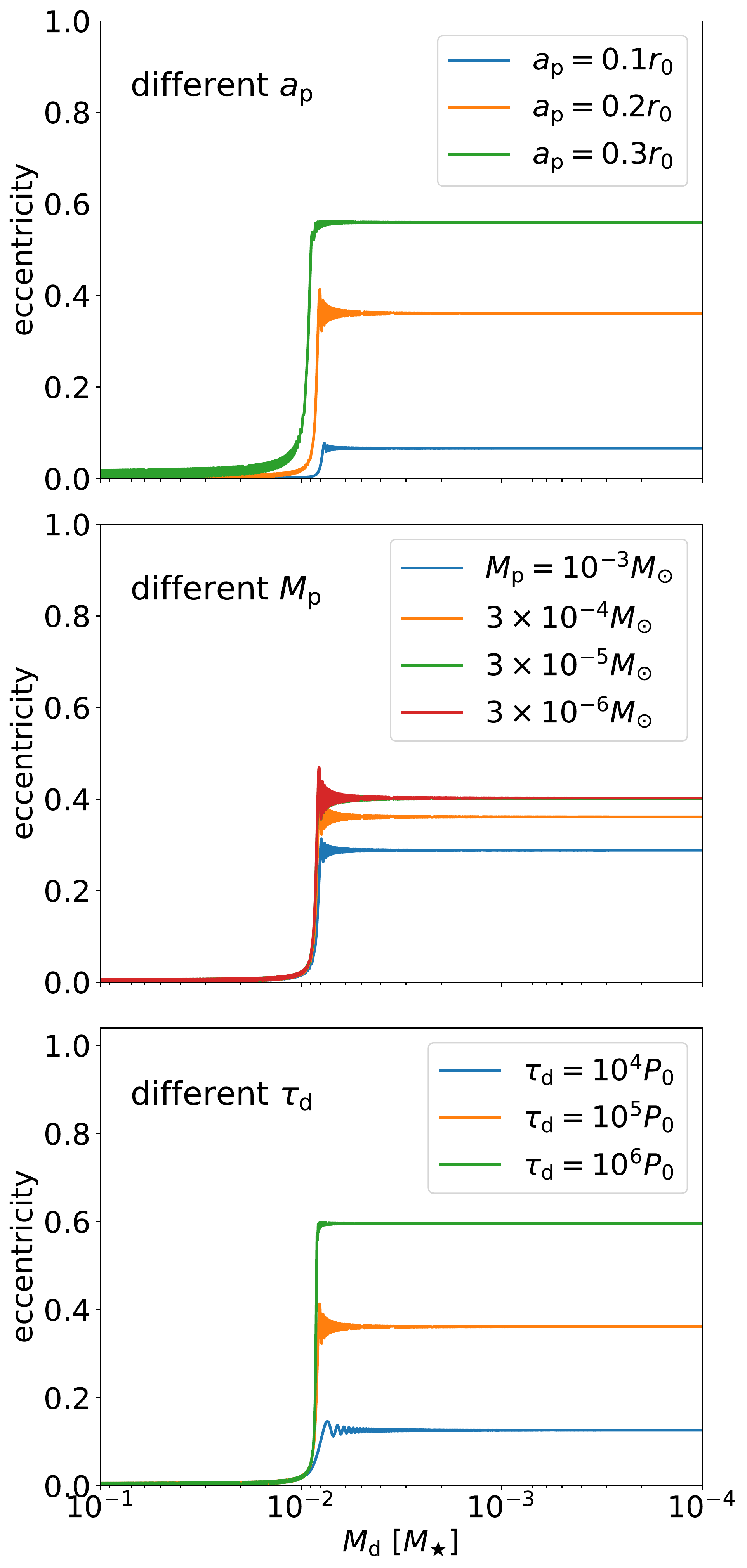}
    \caption{Time evolution of the planet eccentricity in our parameter studies. From top to bottom, each panel shows the results for systems that are the same as the fiducial (see Figure~\ref{fig:res-fid}) but with different planet semi-major axis $a_{\rm p}$, planet mass $M_{\rm p}$, and disk decay timescale $\tau_{\rm d}$, respectively.
    The orange curves in all panels are the same as the $E_{\rm p}$ curve in Figure~\ref{fig:res-fid}.
    }
    \label{fig:res-param}
\end{figure}

For parameter studies, we consider three groups of systems that are the same as the fiducial but with different planet semi-major axis $a_{\rm p}$, planet mass $M_{\rm p}$, and disk decay timescale $\tau_{\rm d}$. 
The results are shown in Figure~\ref{fig:res-param}.
The top panel shows that the final $E_{\rm p}$ is higher when $a_{\rm p}$ is closer to the disk edge $r_0$ because of the stronger planet-disk coupling.

The middle panel of Figure~\ref{fig:res-param} compares the evolution of $E_{\rm p}$ for planets with different masses.
When planets are torqued by the disk, a more massive planet exerts a stronger back-reaction to the disk.
Our result shows that the final $E_{\rm p}$ is smaller for larger $M_{\rm p}$. 
This implies that the back-reaction from the planet tends to reduce the eccentricity transfer.
In the low-mass limit, planets tend to get a same amount of $E_{\rm p}$ because the torque (per unit $M_{\rm p}$) they receive from the disk are the same.

The bottom panel of Figure~\ref{fig:res-param} shows that $E_{\rm p}$ is higher when $\tau_{\rm d}$ is larger. 
Therefore, a slower $\omega_{\rm d}$ sweeping allows more substantial eccentricity transfer than a faster sweeping process.
In Section~\ref{sec:AMD-mode}, we will discuss the role of $\tau_{\rm d}$ from an analytical prospective.

\section{Analysis}
\label{sec:AMD-mode}

In this Section, we use a mode crossing analysis to explain the results of Section~\ref{sec:result}.
Instead of $E_{\rm d}$ and $E_{\rm p}$, it is useful to consider two different variables:
\begin{eqnarray}
    S_{\rm d} & = & \left(\frac{J_{\rm d}}{2}\right)^{1/2}E_{\rm d},\\
    S_{\rm p} & = & \left(\frac{J_{\rm p}}{2}\right)^{1/2}E_{\rm p}.
\end{eqnarray}
Note that $A_{\rm d}=|S_{\rm d}|^2$ and $A_{\rm p}=|S_{\rm d}|^2$ are the angular momentum deficits (AMDs) of the disk and the planet, respectively \citep[see][]{Teyssandier2019MNRAS}.
Equations~\eqref{eq:dEddt} and~\eqref{eq:dEpdt} can be rewritten for $S_{\rm d}$ and $S_{\rm p}$ as
\begin{eqnarray}
\label{eq:SEOM}
    \frac{d}{dt}
    \begin{pmatrix}
    S_{\rm d} \\
    S_{\rm p}
    \end{pmatrix}
    = i
    \begin{pmatrix}
        \omega + \Delta \omega + i\frac{1}{2\tau_{\rm J}} & -\nu \\
        -\nu & \omega - \Delta \omega
    \end{pmatrix}
    \begin{pmatrix}
        S_{\rm d} \\
        S_{\rm p}
    \end{pmatrix},
\end{eqnarray}
where 
\begin{eqnarray}
    \omega 
    & = &
    \frac{\omega_{\rm d} + \omega_{\rm p,d}}{2},\\
    \Delta \omega
    & = &     
    \frac{\omega_{\rm d} - \omega_{\rm p,d}}{2},\\
    \nonumber
    \nu
    & = & \left(\nu_{\rm p,d}\nu_{\rm d,p}\right)^{1/2} \\
    & = &
    \frac{1}{\sqrt{J_{\rm d}J_{\rm p}}}\int GM_{\rm p}\Sigma K_{2}(r,a_{\rm p}) E_{\rm m} 2\pi r dr, \\
    \frac{1}{\tau_{\rm J}}
    & = &
    -\frac{\dot{J}_{\rm d}}{J_{\rm d}} = -\frac{\dot{M}_{\rm d}}{M_{\rm d}}.
\end{eqnarray}
The term $i(2\tau_{\rm J})^{-1}$ in the central matrix in Equation~\eqref{eq:SEOM} can be neglected when $\tau_{\rm J}^{-1}\ll|\Delta \omega|$.
The matrix is then Hermitian, and the mutual interaction between the disk and the planet conserves of the total AMD ($|S_{\rm d}|^2+|S_{\rm p}|^2$).\footnote{Note that because of the $i(2\tau_{\rm J})^{-1}$ term, the total AMD is not conserved in the long-term evolution if the disk is losing mass. However, since resonance crossing occurs over a short period of time, we can drop the $i(2\tau_{\rm J})^{-1}$ term in the analysis.}
The two eigenvectors of the matrix,
\begin{eqnarray}
\label{eq:Spm}
    \vec{S}_{\pm} = 
    \begin{pmatrix}
    S_{\rm d} \\
    S_{\rm p}
    \end{pmatrix}_{\pm}
    = 
    \begin{pmatrix}
    \Delta \omega \pm \sqrt{(\Delta \omega)^2+\nu^2} \\
    -\nu
    \end{pmatrix},
\end{eqnarray}
describe the AMD eigenstates of the ``planet + disk'' system, while
\begin{eqnarray}
    g_{\pm} = \omega \pm \sqrt{(\Delta \omega)^2 + \nu^2}
\end{eqnarray}
are the eigen-frequencies.

\begin{figure}[t]
    \epsscale{1}
    \plotone{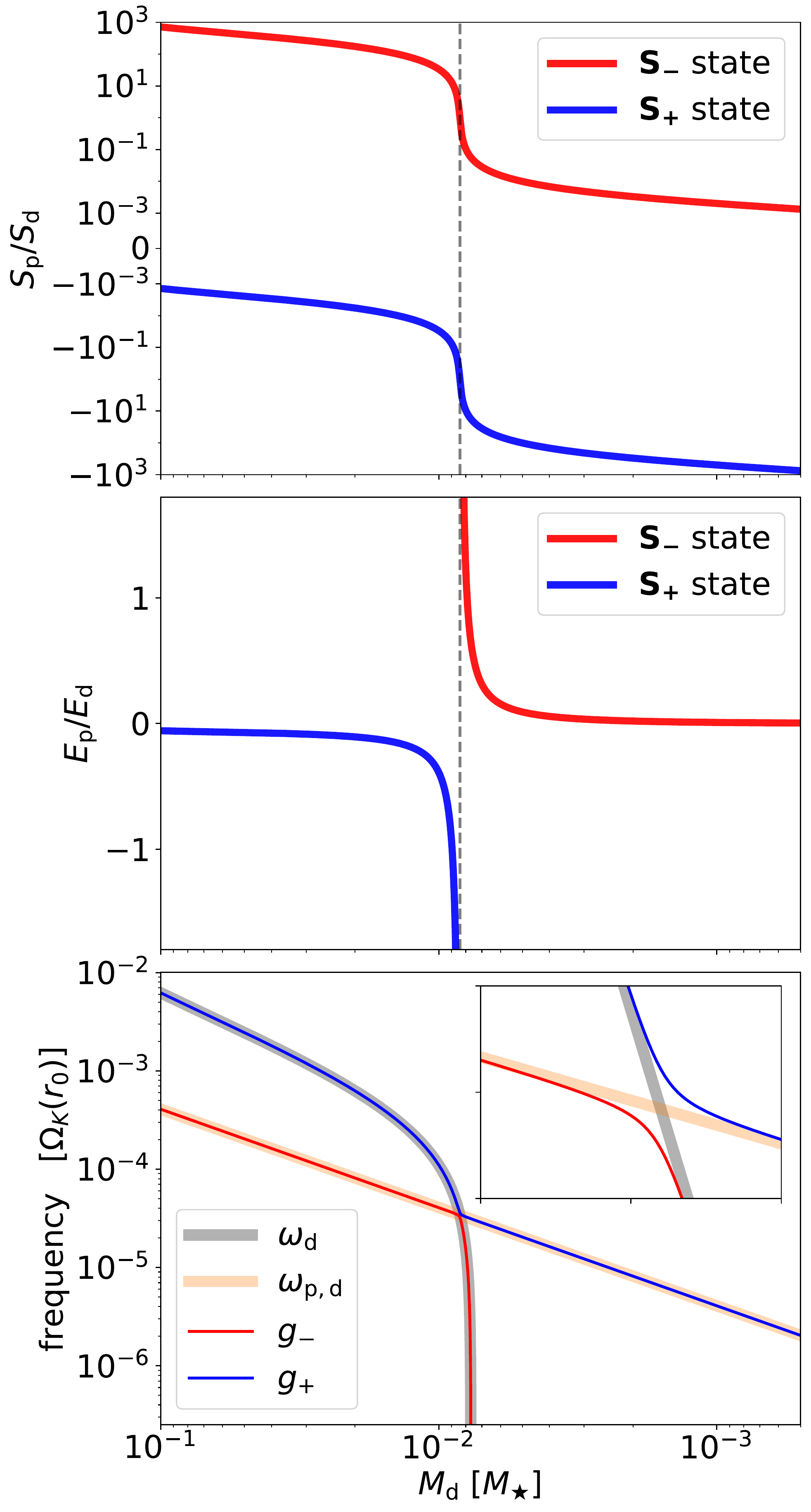}
    \caption{Eigenstates of the ``planet + disk'' system, $\vec{S}_{-}$ and $\vec{S}_{+}$, as functions of the disk mass $M_{\rm d}$. 
    The top panel shows the ratio $S_{\rm p}/S_{\rm d}$ of the two eigenstates. The dashed line marks the $M_{\rm d}$ when $\omega_{\rm d}=\omega_{\rm p,d}$.
    The middle panel shows the corresponding eccentricity ratio $E_{\rm p}/E_{\rm d}$.
    The bottom panel shows the eigenvalues $g_{-}$ and $g_{+}$.
    The values of $\omega_{\rm d}$ and $\omega_{\rm p,d}$ are also shown for comparison. 
    The inset plot zooms in to where $M_{\rm d}$ is between $8\times10^{-3}M_{\star}$ and $9\times10^{-3}M_{\star}$.
    It covers the region around the avoided crossing, which is where $g_{-}$ and $g_{+}$ are closest to each other.
    }
    \label{fig:AMD-state}
\end{figure}

Figure~\ref{fig:AMD-state} plots $\vec{S}_{\pm}$ and $g_{\pm}$ as functions of the disk mass $M_{\rm d}$ for the fiducial system in Section~\ref{sec:result}.
In the $\vec{S}_{-}$ state, the disk and the planet are apsidally aligned (since $S_{\rm p}/S_{\rm d}>0$), and the planet AMD  
$|S_{\rm p}|^2$ is much greater than $|S_{\rm d}|^2$ when the disk is massive and much smaller when the disk is light.
In the $\vec{S}_{+}$ state, the disk and the planet are apsidally anti-aligned ($S_{\rm p}/S_{\rm d}<0$), $|S_{\rm d}|^2\gg|S_{\rm p}|^2$ for large $M_{\rm d}$ and $|S_{\rm d}|^2<|S_{\rm p}|^2$ for small $M_{\rm d}$.
The eigenvalues $g_{\pm}$ are equal to $\omega_{\rm d}$ and $\omega_{\rm p,d}$ for most values of $M_{\rm d}$, except at the resonance $\omega_{\rm d} = \omega_{\rm p,d}$, where $g_{\pm}$ undergo an avoided crossing, with $|g_{+}-g_{-}|$ reaching its minimum value $2\nu$. 

Because the two eigenstates do not cross, if the disk dispersal is slow enough, the time evolution of the ``planet + disk'' system can be solved using the adiabatic invariant: a system in $\vec{S}_{+}$ will follow the blue track in the upper panel of Figure~\ref{fig:AMD-state} as $M_{\rm d}$ decreases, gradually transferring AMD from the disk to the planet;
a system in $\vec{S}_{-}$ will follow the red track with the planet giving AMD to the disk. 

\begin{figure}[t]
    \epsscale{1}
    \plotone{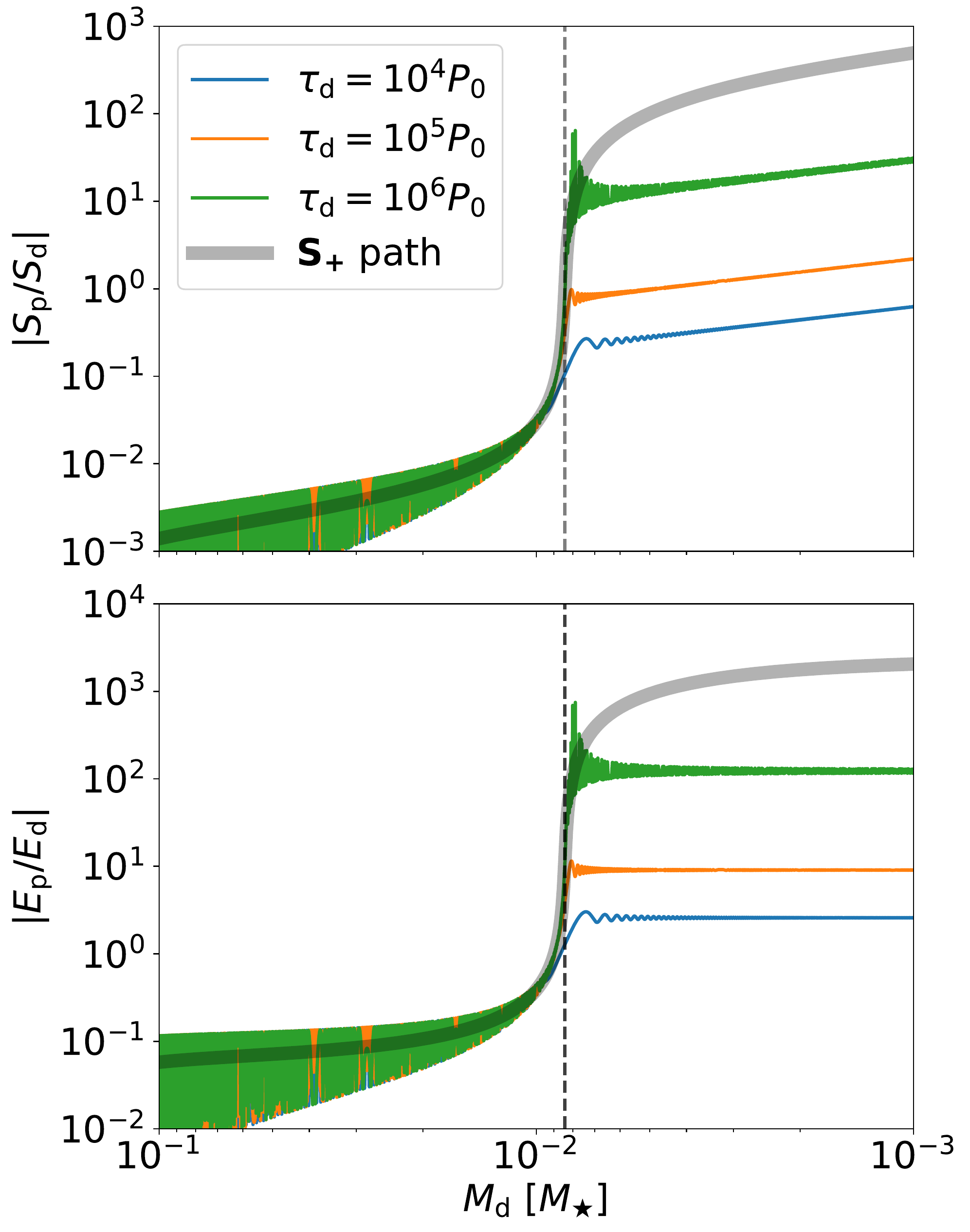}
    \caption{Planet-disk eccentricity evolution for disks with different $\tau_{\rm d}$ (the same runs as shown in the bottom panel of Figure~\ref{fig:res-param}).
    The upper and lower panel show ratio $|S_{\rm p}/S_{\rm d}|$ and $|E_{\rm p}/E_{\rm d}|$, respectively. 
    The grey curve in each panel shows the path of the eccentricity eigenmode $\vec{S}_{+}$ for reference.
    }
    \label{fig:AMD-tau}
\end{figure}

The time evolutions in Section~\ref{sec:result} start with an initial condition that is nearly in the $\vec{S}_{+}$ state.
The upper panel of Figure~\ref{fig:AMD-tau} compares the evolution results for three different $\tau_{\rm d}$'s (from the bottom panel of Figure~\ref{fig:res-param}) with the $\vec{S}_{+}$ trajectory (from the upper panel of Figure~\ref{fig:AMD-state}).
In all three cases, when $M_{\rm d}\gtrsim10^{-2}M_{\star}$, the evolves around the $\vec{S}_{+}$ track.
When $M_{\rm d}\sim10^{-2}M_{\star}$, the eigenstate enters the avoided crossing region and the adiabatic approximation becomes less accurate, especially for the small $\tau_{\rm d}$ case. 
The time evolution trajectory leaves the eigen-track at $|S_{\rm p}/S_{\rm d}|\sim0.1$, $1$, and $10$ for disks with $\tau_{\rm d}=10^4P_0$, $10^5P_0$, and $10^6P_0$, respectively. 
Finally, when $M_{\rm d}\lesssim10^{-2}M_{\star}$, all $|S_{\rm p}/S_{\rm d}|$'s continue to increases because $S_{\rm d}$ decreases with the disk mass.
The lower panel of Figure~\ref{fig:AMD-tau} is the same as the upper, but it compares the eccentricity ratios $|E_{\rm p}/E_{\rm d}|$.

\begin{figure}[t]
    \epsscale{1}
    \plotone{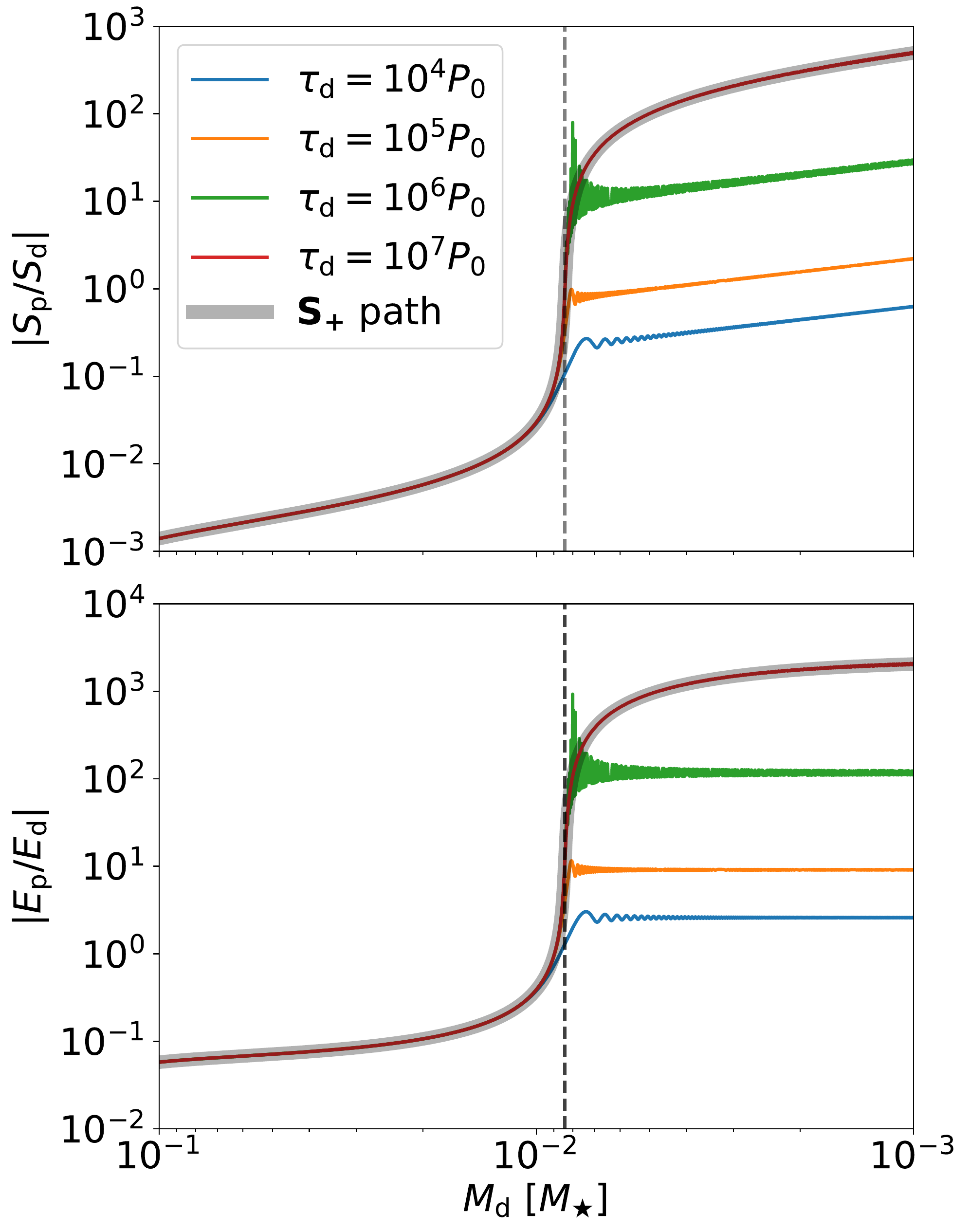}
    \caption{Same as Figure~\ref{fig:AMD-tau}, but shows the evolution of the systems with a slightly non-zero initial planet $E_{\rm p}$ such that the initial condition is exactly in the eccentricity eigenstate $\vec{S}_{+}$.
    }
    \label{fig:AMD-tau-exact}
\end{figure}

Figure~\ref{fig:AMD-tau-exact} is the same as Figure~\ref{fig:AMD-tau}, but shows the evolution of the systems that start exactly on the $\vec{S}_{+}$ track (i.e., the planet is given a slightly non-zero initial eccentricities).
The system now precisely follows the eigen-track until the adiabatic approximation breaks down.
The final $|S_{\rm p}/S_{\rm d}|$'s are the same as the fiducial results.

The condition for adiabatic mode evolution can be evaluated analytically \citep{Lai2002ApJ}.
Define the mode ``mixing angle'' $\theta_{\rm m}$ via
\begin{eqnarray}
    \tan{\theta_{\rm m}} = \frac{\nu}{\Delta\omega+\sqrt{\Delta\omega^2+\nu^2}},
\end{eqnarray}
or, equivalently, via 
\begin{eqnarray}
    \tan{2\theta_{\rm m}} = \frac{\nu}{\Delta\omega},
\end{eqnarray}
then the two eigenmodes (Equation~\ref{eq:Spm}) can be written as 
\begin{eqnarray}
    \vec{S}_{+} = 
    \begin{pmatrix}
    \cos{\theta_{\rm m}} \\
    -\sin{\theta_{\rm m}}
    \end{pmatrix}, \quad
    \vec{S}_{-} = 
    \begin{pmatrix}
    \sin{\theta_{\rm m}} \\
    \cos{\theta_{\rm m}}
    \end{pmatrix}.    
\end{eqnarray}
Adiabatic evolution requires
\begin{eqnarray}
\label{eq:adi-limit}
    \frac{1}{2}\left|g_{+}-g_{-}\right| \gtrsim \left|\frac{d\theta_{\rm m}}{dt}\right|
\end{eqnarray}
during the entirety of the evolution.
The most likely place for this condition to break down is at the resonance (``avoided crossing''), where $\Delta\omega=0$, $\left|g_{+}-g_{-}\right|=2\nu$, $\theta_{\rm m}=\pi/4$, and
\begin{eqnarray}
\label{eq:dthetadt}
    \left|\frac{d\theta_m}{dt}\right|_{\Delta\omega=0} = \left|\frac{1}{2\nu}\frac{d\Delta\omega}{dt}\right|_{\Delta\omega=0}.
\end{eqnarray}
To evaluate $d\Delta\omega/dt$, we use the approximation $|d\Delta\omega/dt| \sim |d\omega_{\rm d}/dt| \sim |8\omega_{\rm d}/\tau_{\rm d}|$ at the resonance crossing (see Figure~\ref{fig:res-fid}).
The adiabaticity condition (Equation~\ref{eq:adi-limit}) therefore becomes
\begin{eqnarray}
    \tau_{\rm d} \gtrsim \tau_{\rm d,crit} \equiv \left|\frac{4\omega_{\rm d}}{\nu^2}\right|_{\Delta\omega=0} = \left(\frac{4\omega_{\rm p,d}}{\nu_{\rm p,d}\nu_{\rm d,p}}\right)_{\Delta\omega=0}.
\end{eqnarray}
Using the approximate expressions~\eqref{eq:w-app} and~\eqref{eq:v-app}, we find 
\begin{eqnarray}
\label{eq:tau-crit}
    \tau_{\rm d,crit} \sim \frac{M_{\star}}{2M_{\rm p}}\left(\frac{r_0}{a_{\rm p}}\right)^4 P_0,
\end{eqnarray}
where $P_0$ is the Keplerian period at $r=r_0$.
For our fiducial parameters ($M_{\rm p}=3\times10^{-4}M_{\star}$, $a_{\rm p}=0.2r_0$), this gives $\tau_{\rm d,crit} \sim 10^6P_0$.  Using more precise values for $\omega_{\rm p,d}$, $\nu_{\rm p,d}$ and $\nu_{\rm d,p}$ gives $\tau_{\rm d,crit} \sim 5\times10^6P_0$.

In Figure 7, we show the result of time integration for a system with $\tau_{\rm d}=10^7P_0$. We see that the planet-disk eccentricity evolution completely traces the eigen-track.

Overall, the above analysis demonstrates that the eccentricity evolution of the planet driven by a dispersing disk found in Section~\ref{sec:result} can be understood using the eccentricity eigenmodes of the composite ``planet+disk'' system. 
Adiabatic evolution along the eigen-track naturally leads to a large-$S_{\rm p}$ or high-$E_{\rm p}$ configuration when the system crosses the resonance (``avoided crossing'') due to disk dispersal. 
For a relatively short disk decay timescale (such as $\tau_{\rm d}\sim 10^5P_0$), the adiabatic approximation is accurate until the system reaches $|S_{\rm p}/S_{\rm d}|\sim1$ around the avoided crossing. 
Because $M_{\rm disk}\gg M_{\rm p}$, the planet may already have received a high eccentricity when the adiabaticity condition no longer holds. 
For a longer $\tau_{\rm d}$, the ``planet + disk'' system can stay on the eigen-track when crossing the resonance, leading to a larger final planet eccentricity.

\section{Conclusion}
\label{sec:conclusion}

In this paper, we have presented a new mechanism of generating large planetary eccentricities.
This mechanism applies to a planet orbiting inside the inner cavity of an initially massive but slowly dispersing protoplanetary disk.
The whole process of eccentricity excitation  includes two stages.

In Stage 1, a massive ($M_d\gtrsim 0.01-0.1M_\star$) protoplanetary disk (with an inner truncation) develops a coherent eccentric structure due to hydrodynamical instability. 
One such instability is the eccentric-mode instability (EMI) \citep{LJR2021ApJ} driven by gas cooling and disk self gravity (Section~\ref{sec:Emode}).
The EMI saturates due to nonlinear hydrodynamical effects, typically at a small disk eccentricity ($E_{\rm d}\sim 0.05$).

In Stage 2, the disk gradually loses mass while preserving a long-lived, precessing eccentric disk mode; the intrinsic mode frequency generally decreases with decreasing disk mass (see Figure~\ref{fig:Emode-w}).
We develop a simple, approximate secular dynamics model \citep[see][]{Teyssandier2019MNRAS} to evolve the eccentricities of the planet and the coherent disk during the disk dispersal  (Section~\ref{sec:planet-disk-interaction}).
We show that the planet can acquire a substantial eccentricity when the ``planet + disk'' system crosses the ``apsidal precession resonance'', when the disk and the planet precesses at the same rate (Section~\ref{sec:result}). 
In our numerical examples (all with an initial disk eccentricity amplitude $E_{\rm d}=0.05$), the final planet eccentricity $e_{\rm p}$ ranges from 0.1 to 0.6, depending on various parameters. 
Higher planetary eccentricities are obtained in systems with larger $a_{\rm p}/r_0$ (the ratio of the planet semi-major axis and the disk truncation radius), smaller planet mass and longer disk dispersal timescale.
We show in Section~\ref{sec:AMD-mode} that the significant eccentricity excitation of the planet by the  dispersing disk can be understood using the eccentricity eigenmodes of the composite ``planet + disk'' system. 
Adiabatic evolution along the eigen-track naturally leads to high-$e_{\rm p}$ when the system crosses the resonance (``avoided crossing'') during the disk dispersal.

We note that the long-term evolution model we have presented in this paper is highly simplified.  
Future hydrodynamical simulations should test the robustness of our model. 
Whether our mechanism occurs depends on disk properties such as the initial disk mass, gas temperature, viscosity, and the mass loss rate.
These properties are still very uncertain, especially for disks that are young and massive (see, e.g., \citealt{Xu2021MNRASa,Xu2021MNRASb} for theoretical modelings of young circum-stellar disks).
Nevertheless, given the recent successes in the disk velocity observations \citep[e.g.,][]{Oberg2021ApJS}, we expect that direct measurement of disk eccentricity profiles will be available in the future.
These data will provide useful constraints on the robustness and efficiency of our mechanism.

\newpage
\acknowledgments
This work is supported in part by NSF grant AST-2107796 and the NASA grant 80NSSC19K0444.

\software{Matplotlib \citep{Hunter2007}, 
NumPy \citep{Walt2011},
SciPy \citep{Virtanen2020}
}


\end{document}